\documentstyle[sprocl,psfig,here]{article}

\def\Journal#1#2#3#4{{#1} {\bf #2}, #3 (#4)}
\def\NPB{{\em Nucl. Phys.} B}
\def\PLB{{\em Phys. Lett.}  B}
\def\PRL{\em Phys. Rev. Lett.}
\def\PRD{{\em Phys. Rev.} D}

\def\be{\begin{equation}}
\def\ee{\end{equation}}
\def\bea{\begin{eqnarray}}
\def\eea{\end{eqnarray}}

\begin{document}
\begin{flushright}
HUPD-9804\\
FERMILAB-Conf-98/022-T\\
hep-ph/9801370\\
January 98
\end{flushright}
\title{TOP PAIR PRODUCTION IN $e^+ e^-$ AND $\gamma\gamma$ PROCESSES}
\author{MICHIHIRO HORI, YUICHIRO KIYO, JIRO KODAIRA, 
        TAKASHI NASUNO \footnote[1]{Talk presented by T. NASUNO 
          at the International Symposium on QCD and 
          New Physics, HIROSHIMA, 1997.}}
\address{Dept. of Physics, Hiroshima University, 
	  Higashi-Hiroshima 739-8526, JAPAN}
\author{STEPHEN PARKE}
\address{Theoretical Physics Department,
	 Fermi National Accelerator Laboratory\\
	 P. O. Box 500, Batavia, IL 60510, USA}
\maketitle
\abstracts{
	We analyze spin correlations between top quark and anti-top
	quark produced at polarized $e^+ e^-$ and $\gamma \gamma$
	colliders.
	We consider a generic spin basis to find a strong spin
	correlation.
	Optimal spin decompositions for top quark pair are presented
	for $e^+ e^-$ and $\gamma \gamma$ colliders.
	We show the cross-section in these bases and discuss the
	characteristics of results.
  	}
\section{Introduction}
In 1994, the top quark was discovered by the CDF and D0
collaborations \cite{CDF,D0}.
The measured top quark mass is approximately $175$ GeV,
which is nearly twice the mass scale of electro-weak symmetry
breaking.
So the top quark with large mass brings a good opportunity to
understand electro-weak symmetry breaking and to search hints of any
new physics.
The top quark decays electroweakly before hadronizing because
its width is much greater than the hadronization timescale set
by $\Lambda_{QCD}$ \cite{BIGI}.
Therefore, there are significant angular correlations between the
decay products of the top quark and the spin of the top quark.
These angular correlations depend sensitively on the top quark
couplings to the Z boson and photon, and to the W boson and b quark.
This aspect is unique for the top quark sector.
Actually, there are many works \cite{HEL} on the angular correlations
for the top quark events produced at $e^+ e^-$ colliders.
In most of these works, the top quark spin is decomposed in the
helicity basis.
However, G. Mahlon and S. Parke \cite{GREG} suggested a more optimal
decomposition of the top quark spin to find spin correlations at hadron
colliders.
S. Parke and Y. Shadmi \cite{PARKE} extended their work to
the $e^+ e^-$ annihilation process at the leading order in perturbation
theory and concluded that the ``Off-Diagonal'' basis is the best
decomposition of top pair spins.

In this paper we focus on the issue of what is the optimal
decomposition of the top quark spins produced at $e^+ e^-$ and $\gamma
\gamma$ colliders \cite{NLC,PLC}.
At $e^+ e^-$ colliders, we analyze the differential cross-sections
for the top quark pair production including QCD one-loop corrections
in the soft gluon approximation.
We give the differential cross-sections in a generic spin basis.
The optimal spin decomposition for the top quark pair and 
the differential cross-section in this basis are presented.
At $\gamma \gamma$ colliders, we discuss the top pair production
with the circular polarized photon beams which correspond to the total
angular momentum $J_{z}=0,2$.
We develop analogous  analyses to the case of the $e^+ e^-$ colliders
and propose an useful spin basis to investigate the spin correlations.
 
%-------------------------------------------------------
\section{Spin Correlations at $e^+ e^-$ Colliders }
\begin{figure}[H]
\begin{center}
\begin{tabular}{cc}
        \leavevmode\psfig{file=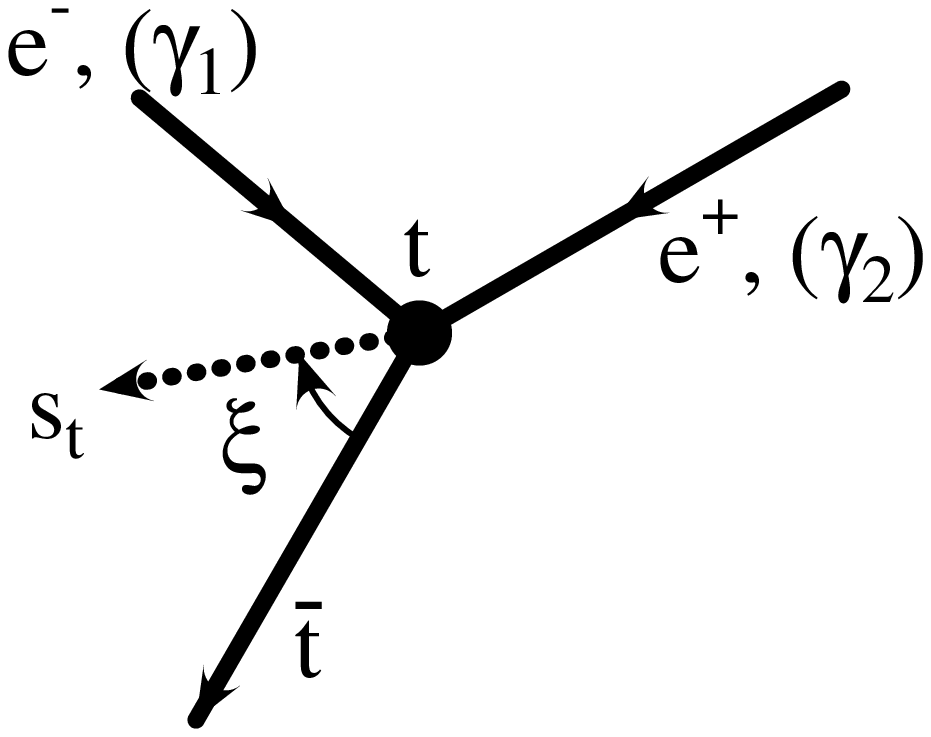,width=4cm} &
        \leavevmode\psfig{file=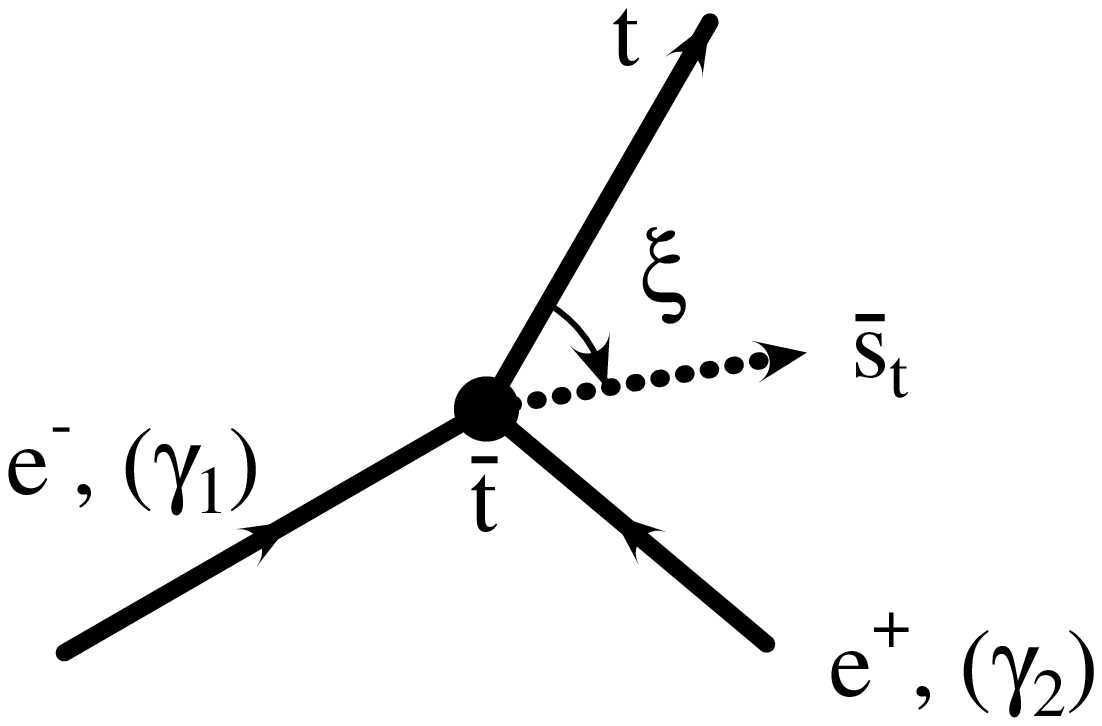,width=4cm} 
\end{tabular}
\caption{The generic spin for the top (anti-top) quark in the top
(anti-top) quark rest frame. 
$\vec{s}_{t}$ ($\vec{s}_{\bar{t}}$) represents the 
spin axis of the top (anti-top) quark.}
\label{fig:SPIN}
\end{center}
\end{figure}

%-------------------------------------------------------
We present the cross-section for the polarized top quark pair
production including the QCD one-loop correction in the soft gluon
approximation.
We show the differential cross-section for a generic spin basis.
The generic spin basis here is based on the following features:
Firstly, we do not discuss the transeverse polarization of the top
quark because it may be much smaller than the longitudinal
polarization.
Secondly CP is conserved.
Therefore, we have defined our generic spin basis so that the spins of 
the top and anti-top quark are in the production plane.
Then, we introduce only one parameter $\xi$ to define the spins
of top and anti-top quarks.
The definition of $\xi$ is as follows [Figure \ref{fig:SPIN}]:
We decompose the top quark spin along the direction $\vec{s}_{t}$ in
the rest frame of the top quark, where it has a relative angle
$\xi$ clock-wisely to the anti-top quark momentum.
Similarly the anti-top spin is defined  in the anti-top quark rest
frame along the direction $\vec{s}_{\bar{t}}$, which makes the same
angle $\xi$ with the top quark momentum.
%-------------------------------------------------------------
%
\vspace*{-0.5cm}
\vspace*{-0.3cm}
%-------------------------------------------------------------

The cross-sections in the Center of Mass (CM) frame are the functions
of the scattering angle $\theta$, the top and anti-top quark speed
$\beta$ and the angle $\xi$ which defines the spin axis. 
In the soft gluon approximation, the differential cross-sections
at the QCD one-loop level are given by
%-------------------------------------------------------
\bea
& &
	\frac{d \sigma }{d \cos \theta  }
	(e^{-}_{L} e^{+}_{R} \rightarrow t_{\uparrow} \bar{t}_{\uparrow}
				~\mbox{and}~
	   			 t_{\downarrow} \bar{t}_{\downarrow})
\nonumber \\
& = &
	\left( \frac{3 \pi \alpha \beta}{2 s} \right)
	(A_{LR} \cos \xi - B_{LR} \sin \xi)
\nonumber \\ && \qquad \qquad  \times
	\bigl[
	(A_{LR} \cos \xi - B_{LR} \sin \xi)(1 + \hat{\alpha}_{s} S_{I})
\nonumber \\ && \qquad \qquad \qquad \qquad \qquad 
      -	2 (\gamma^{2} A_{LR} \cos \xi - \bar{B}_{LR} \sin \xi)
        \hat{\alpha}_{s} S_{II}  
	\bigr]
\label{eqn:1} \\
& &
\frac{d \sigma }{d \cos \theta }
(e^{-}_{L} e^{+}_{R} \rightarrow t_{\uparrow} \bar{t}_{\downarrow}
				~\mbox{or}~
				 t_{\downarrow} \bar{t}_{\uparrow})
\nonumber \\
& = &
	\left( \frac{3 \pi \alpha \beta}{2 s} \right)
	(A_{LR} \sin \xi + B_{LR} \cos \xi \pm D_{LR})
\nonumber \\ && \qquad \qquad \times
	\bigl[
	(A_{LR} \sin \xi + B_{LR} \cos \xi \pm D_{LR})
(1 + \hat{\alpha}_{s} S_{I})
\nonumber \\ && \qquad \qquad \qquad  \qquad \qquad  
      -	2 (\gamma^{2} A_{LR} \sin \xi + \bar{B}_{LR} \cos \xi
           \pm \bar{D}_{LR})
        \hat{\alpha}_{s} S_{II}  
	\bigr]
\label{eqn:2}
\eea
%-------------------------------------------------------
The state $t_{\uparrow}/t_{\downarrow}$ refers to the top quark with
the spin in the direction $+\vec{s}_{t}/-\vec{s}_{t}$ respectively.
Here $\alpha$ is the QED fine structure constant, 
$\hat{\alpha}_{s} = (C_{2}(R)/4 \pi) \alpha_{s}$,
$C_{2}(R) = (N_{c}^{2} - 1)/(2 N_{c})$, $N_{c}$ is the number of
color.
The kinematical variables are $\beta = \sqrt{1 - 4m^2/s}$, 
$\gamma = 1/\sqrt{1 - \beta^{2}}$.
The quantity $A$, $B$, $D$, $\bar{B}$ and $\bar{D}$ are as follows; 
%-------------------------------------------------------
\bea
A_{LR} & = &
[(f_{LL} + f_{LR}) \sqrt{1 - \beta^{2}} \sin \theta ]/2,
\nonumber \\
B_{LR} & = &
[f_{LL}(\cos \theta + \beta ) + f_{LR}(\cos \theta - \beta ) ]/2,
\nonumber \\
D_{LR} & = &
[f_{LL}( 1 + \beta \cos \theta ) + f_{LR}( 1 - \beta \cos \theta ) ]/2,
\label{eqn:3}\\
\bar{B}_{LR} & = &
[f_{LL}(\cos \theta - \beta ) + f_{LR}(\cos \theta + \beta ) ]/2,
\nonumber \\
\bar{D}_{LR} & = &
[f_{LL}( 1 - \beta \cos \theta ) + f_{LR}( 1 + \beta \cos \theta ) ]/2,
\nonumber 
\eea
%-------------------------------------------------------
with
\bea
f_{IJ} ~ = ~
Q_{ \gamma }(e) Q_{ \gamma }(t) + Q_{ Z }^{I}(e) Q_{ Z }^{J}(t)
\frac{1}{ \sin^{2} \theta_{W}} \frac{s}{s - M_{Z}^{2}}.
\nonumber  
\eea
%-------------------------------------------------------
Where $\sqrt{s}$ is the CM energy, $M_{Z}$ is the mass of Z boson,
the angle $\theta_{W}$ is the
Weinberg angle, and the suffix $I,J \in (L,R)$.
The electron (top quark) coupling to the photon is $Q_{\gamma}(e)$
($Q_{\gamma}(t)$).
The left-handed electron (top quark) coupling to the Z boson is given by 
$Q_{Z}^{L}(e)$ ($Q_{Z}^{L}(t)$) and 
the right-handed electron (top quark) coupling to the Z-boson is given by
$Q_{Z}^{R}(e)$ ($Q_{Z}^{R}(t)$).
In the above calculation, we have neglected the Z boson width since
the production threshold for the top quarks is far above the Z boson
mass.
The QCD corrections $S_{I}$ and $S_{II}$ read
%
%-----------------------------------------------------------
\bea
S_{I} & = & 2( \frac{1 +\beta^2}{\beta} \ln \frac{1 + \beta}{1 - \beta} - 2 ) 
\ln \frac{4 \omega_{max}^{2}}{m^2} 
- 8 + 2 \frac{3 + 2 \beta^2}{\beta} 
\ln \frac{1 + \beta}{1 - \beta} 
% +   2 \frac{1 + \beta^2}{\beta}
\nonumber \\
&+& 2 \frac{1 + \beta^2}{\beta}
%& & \times 
[
	2 \zeta(2) + 4 \mbox{\rm Li}_{2} \frac{1 - \beta}{1 + \beta} +
	 \ln \frac{1 - \beta}{1 + \beta} ( 3 \ln \frac{2 \beta}{1 + \beta} + 
                             \ln \frac{2 \beta}{1 - \beta} )
],\quad
\label{eqn:4}
\\
S_{II} 
& = &
\frac{1 - \beta^2}{\beta} \ln \frac{1 + \beta}{1 - \beta},
\label{eqn:5}
\eea
%-----------------------------------------------------------
where $\omega_{max}$ is the upper bound of the soft-gluon energy
and $m$ is the top quark mass.
The factor $S_{I}$ gives the correction  
for the vector (the vector and the axial vector) coupling of the top
quark to the photon (Z boson).
While the factor $S_{II}$ gives the correction of the magnetic moment
of the top quark, $\sigma^{\mu \nu}q_{\nu}$ in the vertices.      
(The momentum $q$ is the incoming momentum of the photon and the Z boson.)
The cross-sections for $e^{-}_{R} e^{+}_{L}$ are given from 
Eqns.(\ref{eqn:1}) - (\ref{eqn:3}) by interchanging $L$, $R$ as well
as $\uparrow$, $\downarrow$.

\begin{figure}[H]
\begin{center}
        \leavevmode\psfig{file=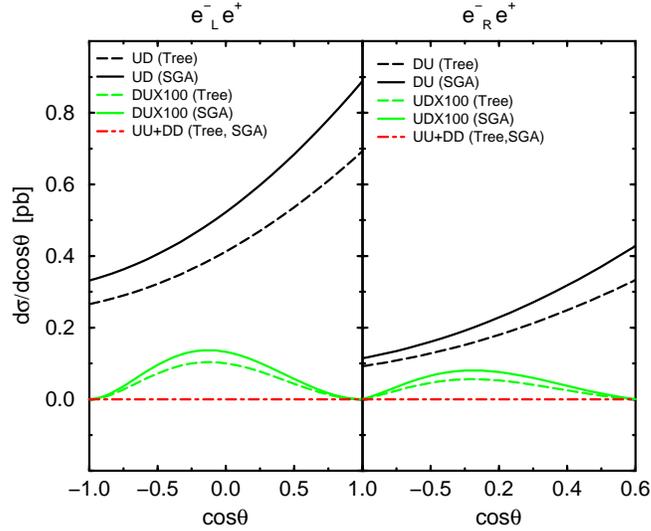,angle=-90,width=9.8cm}
\caption{The differential cross-sections in the Off-Diagonal basis at
 $\sqrt{s}=400$ GeV, $\omega_{max}=10$ GeV for the 
$e^{-} e^+ \rightarrow t \bar{t}$ process:
 $t_{\uparrow} \bar{t}_{\downarrow}$ (UD),
 $t_{\downarrow} \bar{t}_{\uparrow}$ (DU) and 
 $t_{\uparrow} \bar{t}_{\uparrow} 
+ t_{\downarrow} \bar{t}_{\downarrow}$ (UU$+$DD).
The suffix ``Tree'' and ``SGA'' mean the differential cross-section at the
tree level and at the one-loop level in the soft gluon
approximation.
We emphasize that DU (UD) component for the $e^{-}_{L} e^{+}$
($e^{-}_{R} e^{+}$) process is multiplied by 100.} 
\label{fig:fig2}
\end{center} 
\end{figure}

Now we come to the discussion of the spin basis which makes the spin
correlation maximize.
It has been known that there exists the ``Off-Diagonal''
basis which makes the contributions from the like spin configuration
vanish for the $e^-_{L} e^+ \rightarrow t \bar{t}$ process in the
leading order analysis \cite{PARKE}.
At the order $\cal{O}\mit (\alpha_{s})$, we can also employ the Off-Diagonal
basis,
%-----------------------------------------------------------
\bea
\tan \xi = 
\frac{A_{LR}}{B_{LR}}
=
\frac{ (f_{LL} + f_{LR}) \sqrt{1 - \beta^{2}} \sin \theta }
{f_{LL}( \cos \theta + \beta ) + f_{LR}( \cos \theta - \beta )}.
\label{eqn:6}
\eea
%-----------------------------------------------------------
This relation is the same as that in the leading order analysis
\cite{PARKE}.
(The relation for $e^{-}_{R} e^{+}$ is given 
from Eq. (\ref{eqn:6}) by interchange $L$ and $R$.)
The cross-sections are,
%
%-----------------------------------------------------------
\bea
& &  
\frac{d \sigma }{d \cos \theta  }
(e^{-}_{L} e^{+} \rightarrow t_{\uparrow} \bar{t}_{\uparrow}
				~\mbox{and}~
				 t_{\downarrow} \bar{t}_{\downarrow})
= 0, 
\label{eqn:7}
\\
& &
 \frac{d \sigma }{d \cos \theta  }
(e^{-}_{L} e^{+} \rightarrow t_{\uparrow} \bar{t}_{\downarrow}
                                ~ \mbox{or}~      
                                 t_{\downarrow} \bar{t}_{\uparrow})
\nonumber \\
& = &
\left( \frac{3 \pi \alpha^{2}}{4 s \beta} \right)
( \sqrt{A_{LR}^{2} + B_{LR}^{2}} \mp D_{LR} )
\bigl[ 
( \sqrt{A_{LR}^{2} + B_{LR}^{2}} \mp D_{LR} )
        (1 + \hat{ \alpha_{s} }S_{I} ) 
\nonumber \\
& & \qquad\qquad\qquad\qquad\qquad
- 2 (
\frac{ \gamma^2 A_{LR}^2 + B_{LR} \bar{B}_{LR} }
     { \sqrt{A_{LR}^{2} + B_{LR}^{2}}  } \mp
     \bar{D}_{LR})
	 \hat{ \alpha_{s} } S_{II} 
\bigr].
\label{eqn:8}
\eea
%-----------------------------------------------------------
The differential cross-sections in the Off-Diagonal basis are shown in 
Figure \ref{fig:fig2}.
The Up-Up ($t_{\uparrow} \bar{t}_{\uparrow}$) and 
the Down-Down ($t_{\downarrow} \bar{t}_{\downarrow}$) components are
identically zero.
Therefore the total cross-section consists of 
the Up-Down ($t_{\uparrow} \bar{t}_{\downarrow}$) and
the Down-Up ($t_{\downarrow} \bar{t}_{\uparrow}$) components.
But the dominant component is only Up-Down ($t_{\uparrow}
\bar{t}_{\downarrow}$), which makes up more than $99 \%$ of the total
cross-section at $\sqrt{s}=400$ GeV.
Although there exists the structure $\sigma^{\mu \nu} q_{\nu}$ in the
QCD one-loop correction, it does not change the behavior of the
differential cross-section in the Off-Diagonal basis.
The QCD correction makes the differential cross-section
large by $\sim 20 \%$ compared to the tree level cross-section.
Thus the Off-Diagonal basis also gives strong correlation to the
differential cross-section in the soft-gluon approximation.
This means that the Off-Diagonal basis is a really good one.
The hard gluon emission, which we do not discuss, leads to the
spin flip effects.
Therefore it potentially changes the above results.
However the spin flip from the real gluon emission is suppressed by 
$\alpha_{s} \times (E_{g}/M_{t})$, and the amplitude for emitting a hard
gluon is suppressed by the phase space integral.
Here $E_{g}$ is the energy of emitting gluon.
So hard gluon emission may not change the above result.
In fact, the detailed analysis \cite{NEW} suggests that the spin-flip
effects from the hard gluon emission are quite small.
%-----------------------------------------------------------
\vspace*{-0.5cm}
\vspace*{-0.3cm}
%-----------------------------------------------------------
%-----------------------------------------------------------
\section{Spin Correlations at $\gamma \gamma$ Colliders }
At the Next Linear Colliders \cite{NLC}, we have high energy polarized
photon beams which will be produced by the inverse Compton scattering.
Therefore the top quark pair production at $\gamma \gamma$ colliders
is worthy of investigation.
We only discuss the circular polarized photon beams to get clear
information on the top and anti-top quark spins, and 
investigate the cross-section for the process 
($\gamma_{R,L} \gamma_{R,L} 
	\rightarrow 
t_{\uparrow,\downarrow} \bar{t}_{\uparrow,\downarrow} $)
in the case of the total angular momentum $J_{z}=0,2$
(for the detail, see Ref. \cite{GAMMA}).
The suffix $\uparrow$ $(\downarrow)$ denotes the spin up (down) for
the top and the anti-top quarks and the state $\gamma_{R}$ $(\gamma_{L})$
refers to the right-handed (left-handed) photon. 
For the generic spin basis, we take the same definition as
the previous one [Figure \ref{fig:SPIN}].

Firstly, we investigate the optimal decomposition of the top pair spins
for the $J_{z}=0$ channel.
The differential cross-sections for this channel are
%----------------------------------------------------
\bea
\frac{d\sigma}{d\cos\theta}
\left( \gamma_{R}~ \gamma_{R} 
   \rightarrow t_{\uparrow} \bar{t}_{\uparrow} ~\mbox{or}~
               t_{\downarrow}\bar{t}_{\downarrow}
\right)  &=&
y\left( \beta ,\theta \right)
\left(1- \beta^{2} \right) 
\left( 1\mp \beta \cos\xi \right)^{2},\qquad \\
\frac{d\sigma}{d\cos\theta}
\left( \gamma_{R}~ \gamma_{R} 
         \rightarrow t_{\uparrow} \bar{t}_{\downarrow} ~\mbox{and}~
                     t_{\downarrow}\bar{t}_{\uparrow}
\right) &=&
y\left( \beta ,\theta \right)
\left(1-\beta^{2}\right) 
\beta^{2}\sin^{2}\xi.\qquad
\eea
%----------------------------------------------------------
The function $y(\beta,\theta)$ is a common factor, which takes the
following form;
%----------------------------------------------------------
\[
y(\beta,\theta) \equiv
\frac{ \beta }{32 \pi s} %\times
\frac{4 N_{c} (4 \pi \alpha)^2 Q_{\gamma}(t)^4}
     {(1 - \beta^2 \cos^2 \theta)^2}
.
\]
%----------------------------------------------------------
Although we can not take the ``Off-Diagonal'' basis in this case, 
we can employ the ``Diagonal'' basis to correlate the top spins strongly,
This basis is the familiar Helicity basis.
In this basis, we obtain the differential cross-section:
%----------------------------------------------------------
\bea
\frac{d\sigma}{d\cos\theta}
\left(
\gamma_{R}~ \gamma_{R} 
\rightarrow t_{R} \bar{t}_{R} ~\mbox{or}~
t_{L}\bar{t}_{L}
\right)
&=&
y\left( \beta ,\theta \right)
%\times
\left(1-\beta^{2}
\right) 
\left( 1 \pm \beta 
\right)^{2},\qquad\qquad \\
\frac{d\sigma}{d\cos\theta}
\left(
\gamma_{R}~ \gamma_{R} 
\rightarrow t_{R} \bar{t}_{L} ~\mbox{and}~
t_{L}\bar{t}_{R}
\right) 
& = & 0,
\eea
%----------------------------------------------------------
where and in what follows $t_{R/L}$ $(\bar{t}_{R/L})$ refers to the
top (anti-top) with spin up/down in the ``Helicity'' basis 
($\cos \xi = -1$).
%----------------------------------------------------------
\vspace*{-0.5cm}
\begin{center}
\begin{figure}[H]
\leavevmode\psfig{file=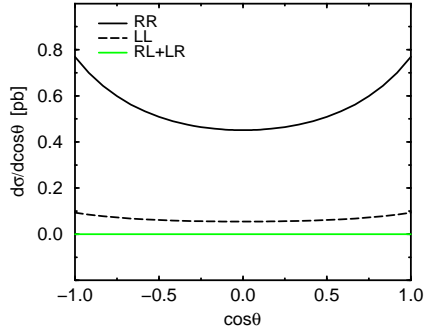,width=6cm}
\caption{The differential cross-sections for the process 
$\gamma_{R} \gamma_{R} \rightarrow t \bar{t}$ at
$\sqrt{s} = 400$ GeV. Each lines corresponds to the spin configuration 
$t_{R} \bar{t}_{R}$, $t_{L} \bar{t}_{L}$ and $t_{R} \bar{t}_{L}+ t_{L}
\bar{t}_{R}$.}
\label{fig:RR}
\end{figure}
\end{center}
\vspace*{-0.3cm}
%-----------------------------------------------------------
In Figure \ref{fig:RR} we show the differential cross-section for the $J_{z}=0$
channel.
The component RR ($t_{R} \bar{t}_{R}$) gives the dominate contribution 
to the total cross-section.
The component LL ($t_{L} \bar{t}_{L}$) is strongly suppressed by the
factor $(1 - \beta)^2$ compared to the $t_{R} \bar{t}_{R}$ at
$\sqrt{s} = 400$ GeV ($\beta \simeq 0.48$).
Roughly speaking, the ratio is 
$d \sigma \left( t_{R} \bar{t}_{R} \right): d \sigma \left
( t_{L}\bar{t}_{L} \right) = 8:1$ .
%-----------------------------------------------------------
%\vspace*{-0.5cm}
%\begin{center}
%\begin{figure}[H]
%\leavevmode\psfig{file=RRdif.ps,width=7cm}
%\caption{The differential cross-sections for the process 
%$\gamma_{R} \gamma_{R} \rightarrow t \bar{t}$ at
%$\sqrt{s} = 400$ GeV. Each lines corresponds to the spin configuration 
%$t_{R} \bar{t}_{R}$, $t_{L} \bar{t}_{L}$ and $t_{R} \bar{t}_{L}+ t_{L}
%\bar{t}_{R}$.}
%\end{figure}
%\end{center}
%-----------------------------------------------------------

\begin{figure}[H]
\begin{center}
\leavevmode\psfig{file=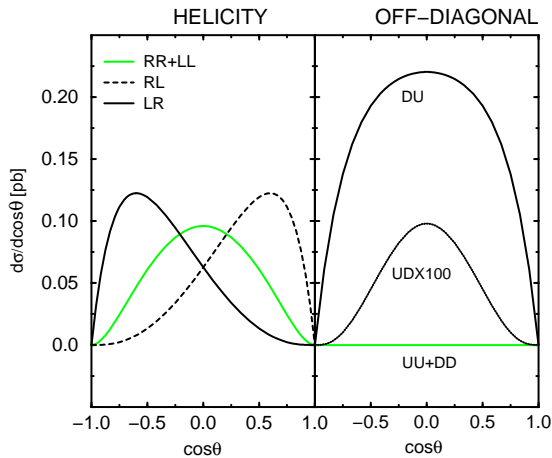,angle=-90,width=8cm} 
\end{center}
\caption{The differential cross-section for the $J_{z}=2$ channel
($\gamma_{R} \gamma_{L}$) in the Helicity and in the Off-Diagonal
bases :RR+LL ($t_{R} \bar{t}_{R} + t_{L}\bar{t}_{L}$), 
RL ($t_{R} \bar{t}_{L}$) and 
LR ($t_{L} \bar{t}_{R}$) in the Helicity basis.
UU+DD ($t_{\uparrow} \bar{t}_{\uparrow} + t_{\downarrow}\bar{t}_{\downarrow}$), 
UD ($t_{\uparrow} \bar{t}_{\downarrow}$) and 
DU ($t_{\downarrow} \bar{t}_{\uparrow}$).
The UD configuration is multiplied by 100.}
\label{fig:RL}
\end{figure}

Secondly, we discuss the spin correlation for the process 
$\gamma_{R} \gamma_{L} \rightarrow t \bar{t}$ in which the initial
angular momentum is $J_{z}=2$. 
The differential cross-sections in generic spin basis are
%----------------------------------------------------------
\bea
&& \frac{d\sigma}{d\cos\theta}
(
\gamma_{R}~ \gamma_{L}
 \rightarrow  t_{\uparrow} \bar{t}_{\uparrow} ~\mbox{and}~
t_{\downarrow} \bar{t}_{\downarrow}
) 
\nonumber \\
& & =
y\left( \beta ,\theta \right)
%\times
\beta^{2} \sin^{2}\theta
\left(
 \sqrt{1 - \beta^2} \sin \theta \cos\xi 
- \cos \theta \sin \xi 
\right)^{2},\qquad \qquad \\
&& \frac{d \sigma}{d\cos\theta}
(
\gamma_{R}~ \gamma_{L} 
\rightarrow t_{\uparrow} \bar{t}_{\downarrow} ~\mbox{or}~
t_{\downarrow}\bar{t}_{\uparrow}
) 
\nonumber \\ 
& & =
y\left( \beta ,\theta \right)
%\times
\beta^{2}\sin^{2}\theta
\left(
 \sqrt{1 - \beta^2} \sin \theta \sin\xi + 
 \cos \theta \cos\xi \mp 1
\right)^{2}.\qquad \qquad 
\eea
%----------------------------------------------------------
In the ``Helicity'' basis with $\cos \xi= -1$, the differential
cross-sections are
%----------------------------------------------------------
\bea
\frac{d\sigma}{d\cos\theta}
(
\gamma_{R}~ \gamma_{L}
 \rightarrow  t_{\uparrow} \bar{t}_{\uparrow} ~\mbox{and}~
t_{\downarrow} \bar{t}_{\downarrow}
) 
& = &
y\left( \beta ,\theta \right)
\beta^{2} \sin^{4}\theta
\left(
1 - \beta^2
\right),\qquad \qquad \\
\frac{d \sigma}{d\cos\theta}
(
\gamma_{R}~ \gamma_{L} 
\rightarrow t_{\uparrow} \bar{t}_{\downarrow} ~\mbox{or}~
t_{\downarrow}\bar{t}_{\uparrow}
) 
& = &
y\left( \beta ,\theta \right)
\beta^{2} \sin^{2} \theta
\left(
 \cos \theta \pm 1
\right)^2.\qquad \qquad 
\eea
%----------------------------------------------------------
Here we can take the ``Off-Diagonal'' basis by defining the spin
angle $\xi$ as follows,
%----------------------------------------------------------
\bea
\tan \xi ~=~ \sqrt{1 - \beta^{2}} \tan \theta.
\eea
%----------------------------------------------------------
In this basis, we get the differential cross-sections, 
%----------------------------------------------------------
\bea
\frac{d \sigma}{d\cos\theta}
\left(
\gamma_{R}~ \gamma_{L} 
\rightarrow t_{\uparrow} \bar{t}_{\uparrow} ~\mbox{and}~
t_{\downarrow}\bar{t}_{\downarrow}
\right) 
&=& 0 ,\\
\frac{d\sigma}{d\cos\theta}
\left(
\gamma_{R}~ \gamma_{L} 
\rightarrow t_{\uparrow} \bar{t}_{\downarrow} ~\mbox{or}~
t_{\downarrow}\bar{t}_{\uparrow}
\right) 
&=&
y\left( \beta ,\theta \right)
\beta^{2} \sin^{2}\theta
\nonumber \\
&& \times
\left(
1 \mp \sqrt{1-\beta^{2}\sin^{2}\theta}
\right)^{2}.
\qquad\qquad
\eea\\
%-----------------------------------------------------------
To see the difference between the Helicity and the Off-Diagonal basis,
we plot the differential cross-section at $\sqrt{s} = 400$ GeV [Figure
\ref{fig:RL}].
In the Helicity basis, the cross-section is not dominated by only one
spin configuration.
While the spin configuration Down-Up ($t_{\downarrow} \bar{t}_{\uparrow}$)
dominates the cross-section in the Off-Diagonal basis.
Therefore we can uniquely determine the spin configuration of the top
quark pairs to be ``Down-Up'' ($t_{\downarrow} \bar{t}_{\uparrow}$) in
the Off-Diagonal basis.
%-----------------------------------------------------------
\vspace*{-0.5cm}
\vspace*{-0.3cm}
%----------------------------------------------------------
\section{Summary}
Firstly, the top quark pair production at $e^+ e^-$ Colliders was
discussed.
We have investigated the spin correlation at the one-loop level in the
soft-gluon approximation and shown the polarized cross section in the
Off-Diagonal basis.
In the Off-Diagonal basis, the QCD correction in the soft-gluon
approximation does not change the behavior of the cross-section.
It only enhances the magnitude of cross-section by $\sim 20\%$
compared to the ones in the leading order analysis.
It has been shown that the Off-Diagonal basis is effective even after taking 
into account the QCD corrections.

Secondly we have discussed the spin correlations at $\gamma \gamma$ Colliders at
the tree level in the perturbation theory.
The characteristic spin structure which makes spin
correlations strong was discussed.
In the $J_{z}=0$ channel, the Helicity basis is a good basis.
The dominant component of the signal is RR ($t_{R} \bar{t}_{R}$),
which makes up about $90 \%$ of the total cross-section at
$\sqrt{s}=400$ GeV.
In the $J_{z}=2$ channel, we presented the polarized cross-section in the
Helicity and Off-Diagonal basis.
In the Off-Diagonal basis, only one spin configuration is appreciably
different from zero for all values of the scattering angle in contrast 
with the Helicity basis.

To observe spin correlations at future $e^+ e^-$ and the $\gamma
\gamma$ Colliders is interesting and may be a good test for the top
quark sector of the Standard Model.
Especially, the analysis using the Off-Diagonal basis will give us a
lot of useful and efficient information.
%------------------------------------------------------------
\section*{Acknowledgments}
%------------------------------------------------------------
T.N would like to thank organisers for their hospitality and inviting
me to interesting conference.
J.K is supported in part by the Monbusho Grant-in-Aid for Scientific
Research No. C-09640364.
The Fermi National Accelerator Laboratory is 
operated by the Universities Research Association, Inc., under contract
DE-AC02-76CHO3000 with the United States of America Department of Energy.
%------------------------------------------------------------

\end{document}